\documentclass[12pt]{iopart}

\usepackage{iopams}
\usepackage{epsfig} 
\usepackage{psfrag}
\usepackage{braket}
\usepackage{color}
\usepackage[usenames,dvipsnames]{xcolor}
\usepackage{textcomp}
\usepackage{graphicx}
\usepackage{latexsym}
\usepackage[lofdepth,lotdepth]{subfig}
\usepackage[numbers,square,sort&compress]{natbib}
\usepackage[derived,binary,squaren,Gray,thickqspace,thinspace]{SIunits}

\usepackage{hyperref}
\hypersetup{dvips,
  pdfauthor={Christopher N. Varney, Karl A. H. Sellin, Qing-Ze Wang,
    Hans Fangohr, and Egor Babaev},
  pdftitle={Hierarchical structure formation in layered superconducting systems with multi-scale inter-vortex interactions},
  colorlinks=true,
  linkcolor=blue,
  citecolor=blue,
  pdfpagemode=UseNone
}

\usepackage{breakurl}

\providecommand*{\deriv}[3][]{%
  \frac{d^{#1}{#2}}{{d #3}^{#1}}}

\begin{document}

\title[Heirarchical structure formation in layered superconducting systems...]{Hierarchical structure formation in layered superconducting systems with multi-scale inter-vortex interactions}
\author{Christopher N. Varney$^{1,2}$, Karl A. H.  Sellin$^{3}$, Qing-Ze
  Wang$^{1,4}$, Hans Fangohr$^{5}$, and Egor Babaev$^{1,3}$}
\address{$^1$Department of Physics, University of Massachusetts, Amherst,
  Massachusetts 01003, USA}
\address{$^2$Department of Physics, University of West Florida,
  Pensacola, Florida 32514, USA}
\address{$^3$Department of Theoretical Physics, The Royal Institute of
  Technology, SE-10691 Stockholm, Sweden}  
\address{$^4$Department of Physics, The Pennsylvania State University,
  University Park, Pennsylvania 16802, USA}
\address{$^5$Engineering and the Environment, University of
  Southampton, Southampton, SO17 1BJ, UK}
\ead{cvarney@uwf.edu}

\begin{abstract}
  We demonstrate formation of hierarchical structures in
  two-dimensional systems with multiple length scales in the
  inter-particle interaction. These include states such as clusters of
  clusters, concentric rings, clusters inside a ring, and stripes in a
  cluster.  We propose to realize such systems in vortex matter (where
  a vortex is mapped onto a particle with multi-scale interactions) in
  layered superconducting systems with varying inter-layer thicknesses
  and different layer materials.
\end{abstract}

\pacs{
  74.25.Uv, % vortex phases
  74.25.Dw, % Phase diagrams - superconductivity
  74.45.+c, % Proximity effects (superconductivity), 
  61.46.Bc  % Structure of clusters
}

%\submitto{\JPCM}
%\maketitle

\section{Introduction}
\label{sec:intro}
Condensed matter physics has long been concerned with explaining
phenomena that result from competing interactions, covering a wide
variety of topics from soft condensed matter systems to magnetism and
ultra-cold atoms (for a recent overview see
Refs.~\cite{O.Reichhardt2010} and \cite{O.Reichhardt2011}). The
richest pattern forming systems are those with several length
scales. For example, structure formation in systems with multi-scale
interactions is highly relevant in hard condensed matter
systems~\cite{Spivak2004,Smorgrav2005,Parameswaran2012}, nuclear
matter~\cite{Ravenhall1983}, and in colloids and other soft condensed
matter systems~\cite{Malescio2003,Glaser2007}.  Alternatively,
nontrivial patterns can arise from the combination of inter-particle
interactions and an external potential~\cite{YLiu2008}.

One of the promising systems for non-trivial structure formation are
superconducting vortices. Research on the magnetic response of
standard type-2 superconductors traditionally typically deals with
structure formation of vortices~\cite{Abrikosov1957}. Although the
interaction between vortices in these materials has a simple
monotonically repulsive form, the vortex matter exhibits a plethora of
interesting phase transitions and structure
formation~\cite{Blatter1994}. Moreover, the vortex states, especially
in the presence of pinning, are critically important for technological
applications of superconductors, where control over vortex matter in
many cases amounts to control of dissipation.

Recently, the possibility of more complicated inter-vortex
interactions in newly discovered systems has attracted much attention
in various contexts: in multi-component
superconductors~\cite{Babaev2005,Moshchalkov2009,Nishio2010,
  Dolocan2005,Moler2005,Curran2011,Geurts2010,O.Reichhardt2010,
  Carlstrom2011prb83, Carlstrom2011prb84,Lin2011,Silaev2011,
  Silaev2012,Dao2011,Garaud2012,Drocco2013,Gutierrez2012},
superfluids~\cite{Liu2008}, vortices in dense nuclear matter in
neutron stars~\cite{Alford2008}, and quantum Hall
systems~\cite{Parameswaran2012}. Recently a problem which attracted
interest was the phase separation in vortex matter with long-range
attractive and short-range repulsive inter-vortex interactions. Such
forces in multi-component superconductors originate in the regime
where there are several superconducting components originating from
different bands. This gives rise to two coherence lengths, $\xi_1$ and
$\xi_2$, and the magnetic field penetration length falls between them:
$\xi_1<\lambda<\xi_2$~\cite{Babaev2005}. Such multi-scale long-range
attractive, short-range repulsive inter-vortex interaction potentials
were derived in a variety of superconducting models: ranging from
multi-component Ginzburg-Landau models with various interband couplings
\cite{Carlstrom2011prb83} to multi-band Eilenberger
models~\cite{Silaev2011,Silaev2012}. This phenomena, recently referred
to as ``type-1.5 superconductivity,''~\cite{Moshchalkov2009} is the
subject of a recent review~\cite{Babaev2012}
% in Ref.~\onlinecite{Babaev2012}
and has stoked considerable experimental interest in pursuing a
realization of such regimes using artificial structures made of
alternating layers of type-1 and type-2 materials.

% In a broader context structure formation in systems with competing
% interaction is a fundamental question with wide physical
% applications~\cite{Seul1995}. The richest pattern forming systems
% are those with several repulsive length scales. For example,
% structure formation in systems with two-scale repulsive interactions
% is highly relevant in such contexts as nuclear
% matter~\cite{Ravenhall1983} and in colloids and other soft condensed
% matter systems~\cite{Malescio2003,Glaser2007}.

\begin{figure}[t]
  \centering
  \includegraphics[height=0.20\textheight]{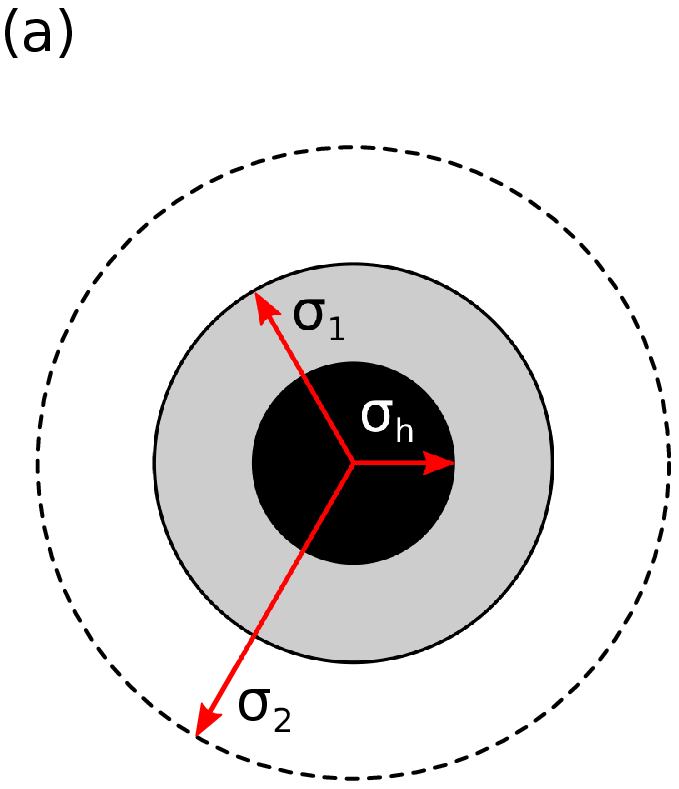}
  \includegraphics[height=0.20\textheight]{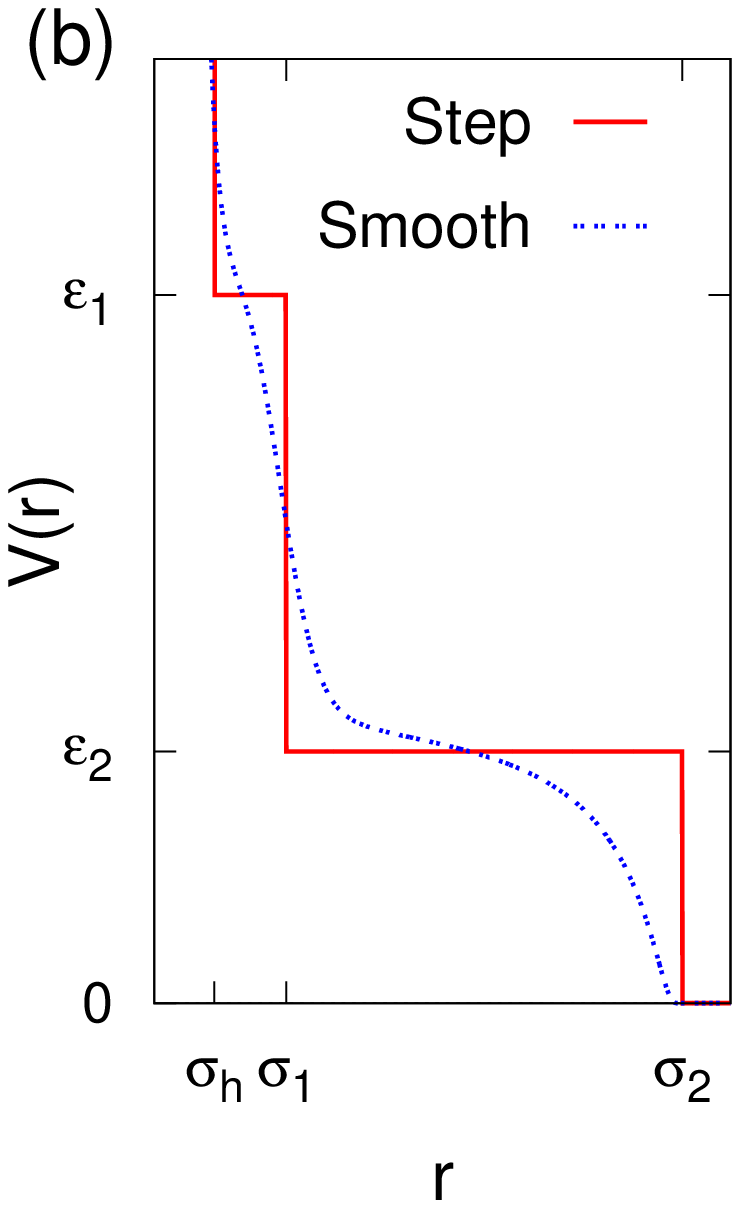}
  \includegraphics[height=0.20\textheight]{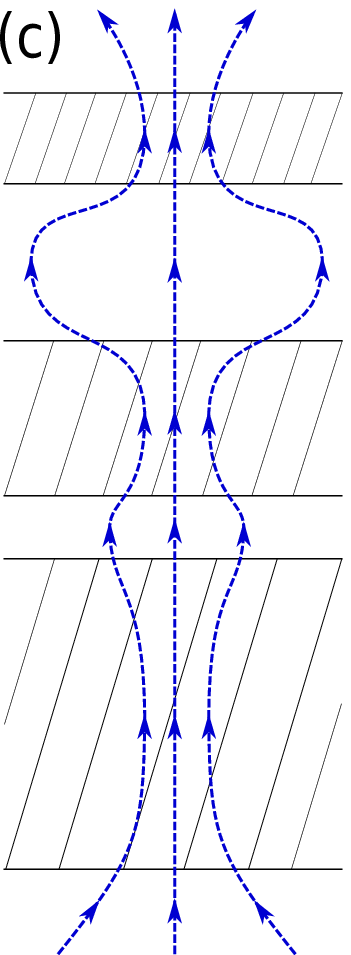}
  \caption{
    (a)~Sketch of a vortex showing interaction length
    scales. $\sigma_h$ is the hard-core radius and $\sigma_1$
    ($\sigma_2$) is the inner (outer) soft-core radius. (b)~Pair
    potential $V(r)$ as a function of the inter-particle separation $r$
    for a step potential and an analogous smooth
    potential. (c)~Schematic drawing of the field lines of a vortex in
    a layered superconductor. Shaded (white) regions are
    superconducting (insulation) of varying thickness.
    Having layers with different magnetic field
    penetration lengths allows for multiple repulsive length
    scales in vortex-vortex interactions.  Also by controlling
    the thickness of the insulating layers one can control the repulsive
    length scales of the inter-vortex interaction. Thicker insulating
    layers cause a wider spread of the magnetic field lines
    resulting in the presence of an additional repulsive length in
    the inter-vortex interaction scales. Having type-I layers also
    allows having repulsive scales in the inter-vortex
    interaction~\cite{Babaev2012}.
    \label{fig:schematic}
  }
\end{figure}

Here we propose that superconducting systems can have vortex states
with several length scales of repulsive (also in some cases
attractive) interactions, where more complicated interaction
potentials can be realized. Such inter-vortex forces should arise in
layered structures made of combinations of type-2 and type-1
superconductors where the magnetic field penetration length $\lambda$
varies in different layers. Since $\lambda$ sets the length scale of
the repulsive inter-vortex forces, having a vortex stack piercing
several layers, each having its own $\lambda_i$ as shown in
\Fref{fig:schematic}(c) should result in the existence of several
repulsive length scales $\lambda_i$ in the interaction between such
vortices (a vortex stack is kept together due to electromagnetic and
inter-layer Josephson coupling). Another way to produce multiple
repulsive length scales is to have different insulating layers of
different thickness. The thickness determines how much magnetic field
lines spread between the layers. The supercurrent on a surface of a
superconductor should be determined self-consistently with the
inter-layer field and thus it should result in multiple repulsive
length scales. The corresponding interaction length scales should be
short range as can be deduced from flux conservation similar to those
in Ref.~\cite{dipolar}. Also, for quasi-two-dimensional systems an
additional power-law repulsive interaction is present due to the
interaction of stray fields outside the sample~\cite{Pearl1964}. Here
we consider the regime where vortex line tension is large and
temperature is small so the vortices do not bend. In this limit we can
describe the system with an effective two-dimensional model where
vortices can be described by their positions in the $xy$-plane. The
results, however, are generalizable to the three-dimensional case
which we will consider separately.

In what follows, we use both Monte Carlo (MC)~\cite{Landau2005} and
Langevin dynamics (LD)~\cite{Fangohr2001} simulations, which are
typically used in studying vortex physics, to demonstrate that vortex
systems with competing interactions ranging over multiple length
scales exhibit nontrivial {\it hierarchical} structure formation,
where at short distances the system can form vortex clusters or vortex
stripes, which subsequently order themselves into complex patterns at
longer length scales. In \Sref{sec:repulsive}, we discuss the ground
state vortex configuration of a hard sphere model with several
repulsive length scales. Next, we examine the more general case of
multi-scale interactions containing regions of repulsion and
attraction in \Sref{sec:gen}. We summarize the conclusions of this
work in \Sref{sec:summary} and conclude with a discussion of the
details of the simulation and the potentials studied in
\ref{sec:methods} and \ref{sec:potentials}, respectively.

\section{Purely Repulsive Interactions}
\label{sec:repulsive}
With the physical realization described in \Sref{sec:intro} in
mind, let us consider the simplest potential with several length
scales, a hard sphere model with multiple shoulders. Note that
``core-softened'' potentials with a single shoulder have been studied
intensively, revealing a myriad of density-modulated ground
states~\cite{Malescio2003,Malescio2004, Camp2003,Camp2005,Glaser2007}.
In Figures~\ref{fig:schematic}(a) and \ref{fig:schematic}(b) we
illustrate a particle with an impenetrable hard-core radius $\sigma_h$
and two repulsive shoulders at $r = \sigma_1$ and $\sigma_2$ with
heights $\varepsilon_1$ and $\varepsilon_2$, respectively.

\begin{figure}[t]
  \centering
  \includegraphics[width=\columnwidth]{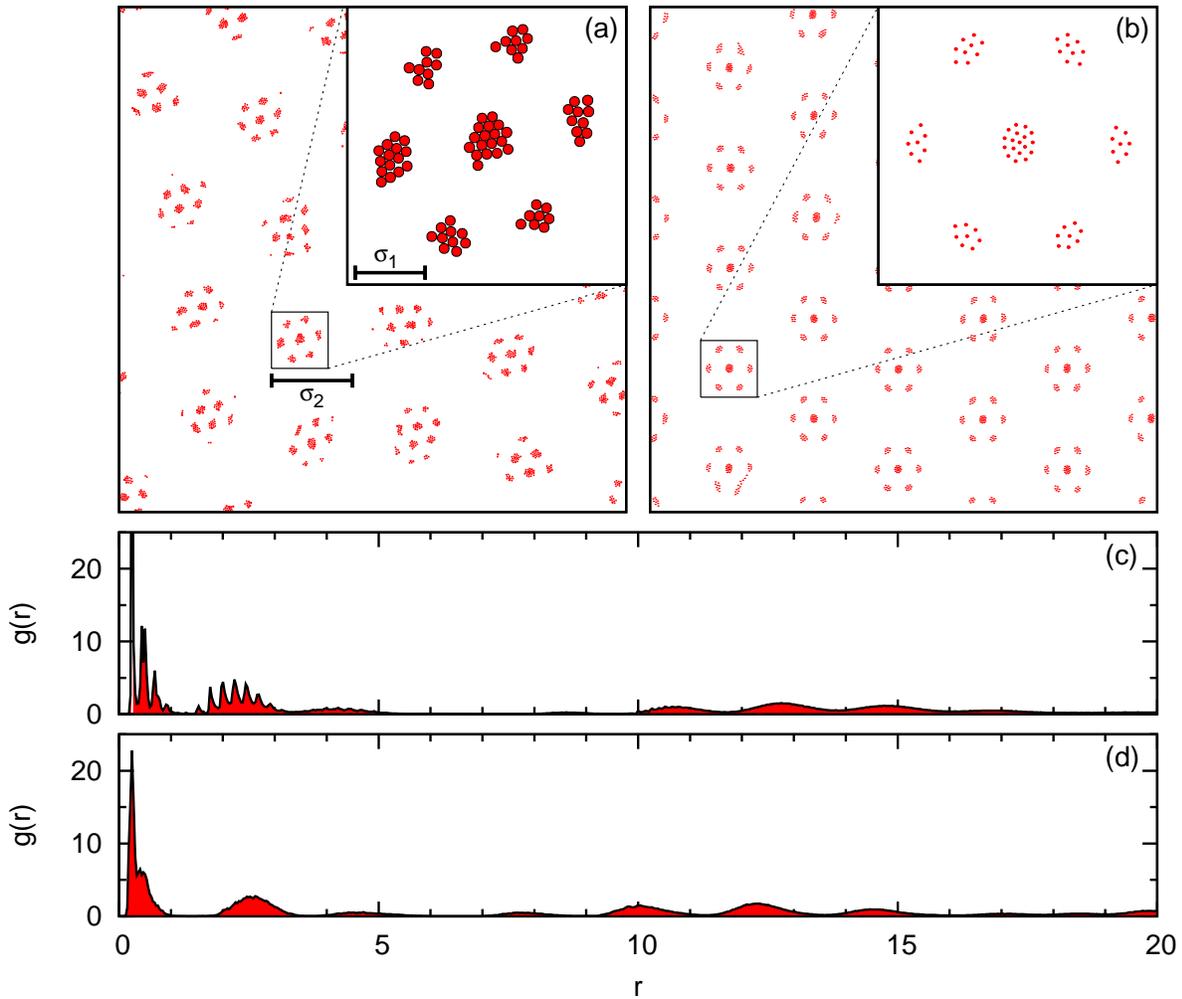}
  \caption{
    Final particle configuration for (a)~three-step
    potential of \Fref{fig:schematic}(b) and (b)~smoothed
    potential with $N_v=2000$ and $\rho=0.51$. The corresponding
    radial distribution functions are plotted versus particle
    separation (in units of the characteristic simulation length
    $\lambda$) in panels (c) and (d), respectively. The length scales
    $\sigma_1$ and $\sigma_2$ are defined in
    \ref{sec:potentials}. 
    \label{fig:step}
  }
\end{figure}

We show the final configuration from MC simulations of our hard-sphere
model in \Fref{fig:step}(a) for $N_v=2000$ particles and density
$\rho=0.51$. Here the system forms a hierarchical
  structure: namely the particles order on three different length
scales: (1)~the particles form a tightly bound cluster, (2)~the
clusters are themselves bound into a conglomerate structure (hereafter
referred to as a supercluster), and (3)~the structures form a
lattice. To analyze the underlying structure of this phase, we show
the radial distribution function (RDF) $g(r)$~\cite{Hansen1986} in
\Fref{fig:step}(c). The first feature in $g(r)$ is a very strong peak
corresponding to the nearest-neighbor distance inside each
cluster. Because the clusters have such a small radius, $g(r)$ shows
only a small peak at approximately double the nearest-neighbor
distance. The next pronounced peak is the inter-cluster distance
inside a supercluster, with the subsequent peaks describing the
distance between clusters in different superclusters.

\begin{figure}[t]
  \centering
  \includegraphics[width=0.8\columnwidth]{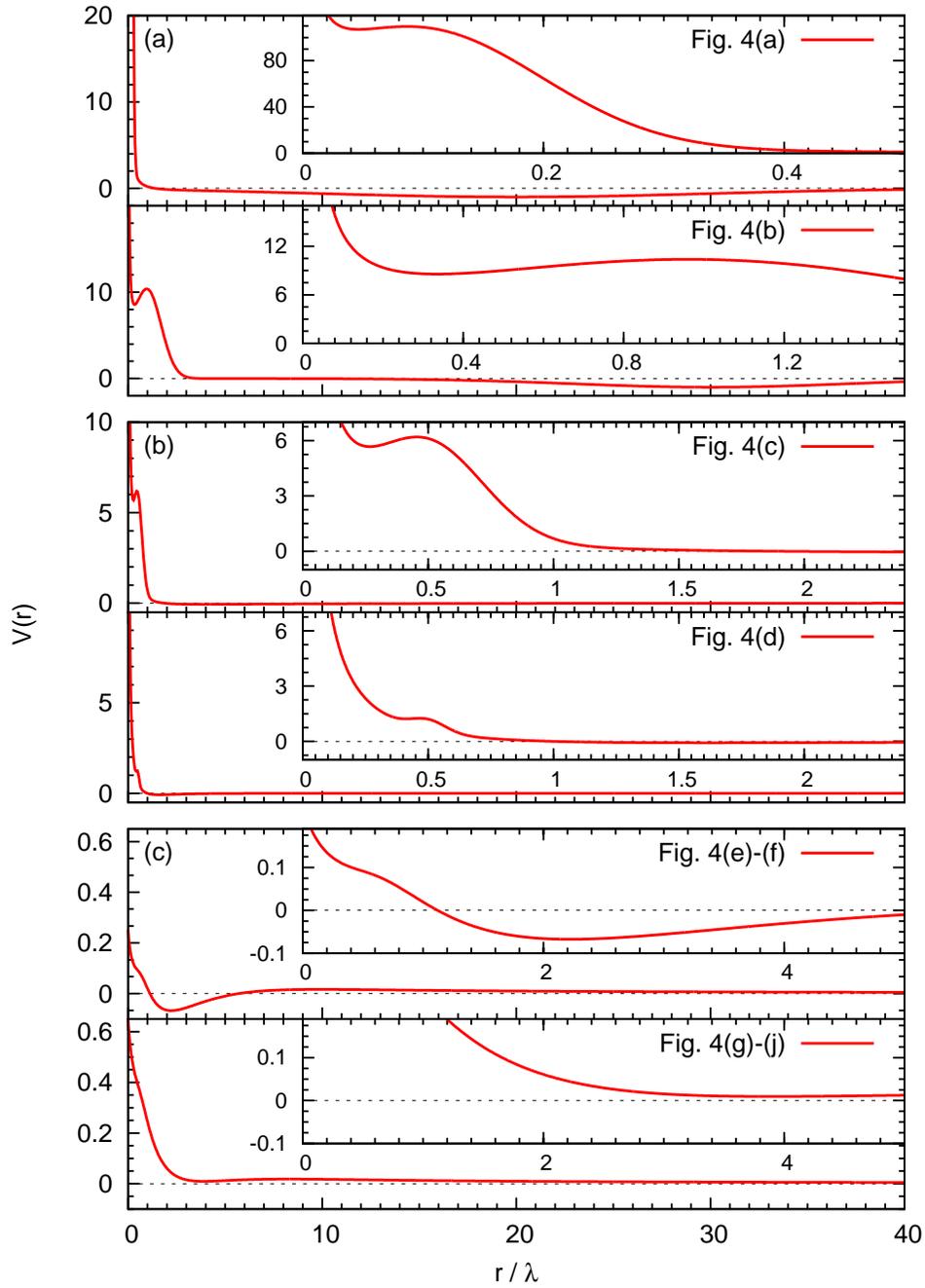}
  \caption{
    Inter-vortex pair potentials used in this
    study. Panels (a), (b), and (c) each illustrate a different pair
    potential, with the top and bottom of each panel representing a
    different set of parameters (see Appendix B for details). The
    insets are close up views of the potentials for small
    inter-particle separation. The legend indicates the corresponding
    panels in \Fref{fig:snapshots}.
    \label{fig:potentials}
  }
\end{figure}

\begin{figure*}[htb]
  \centering
  \includegraphics[width=\textwidth]{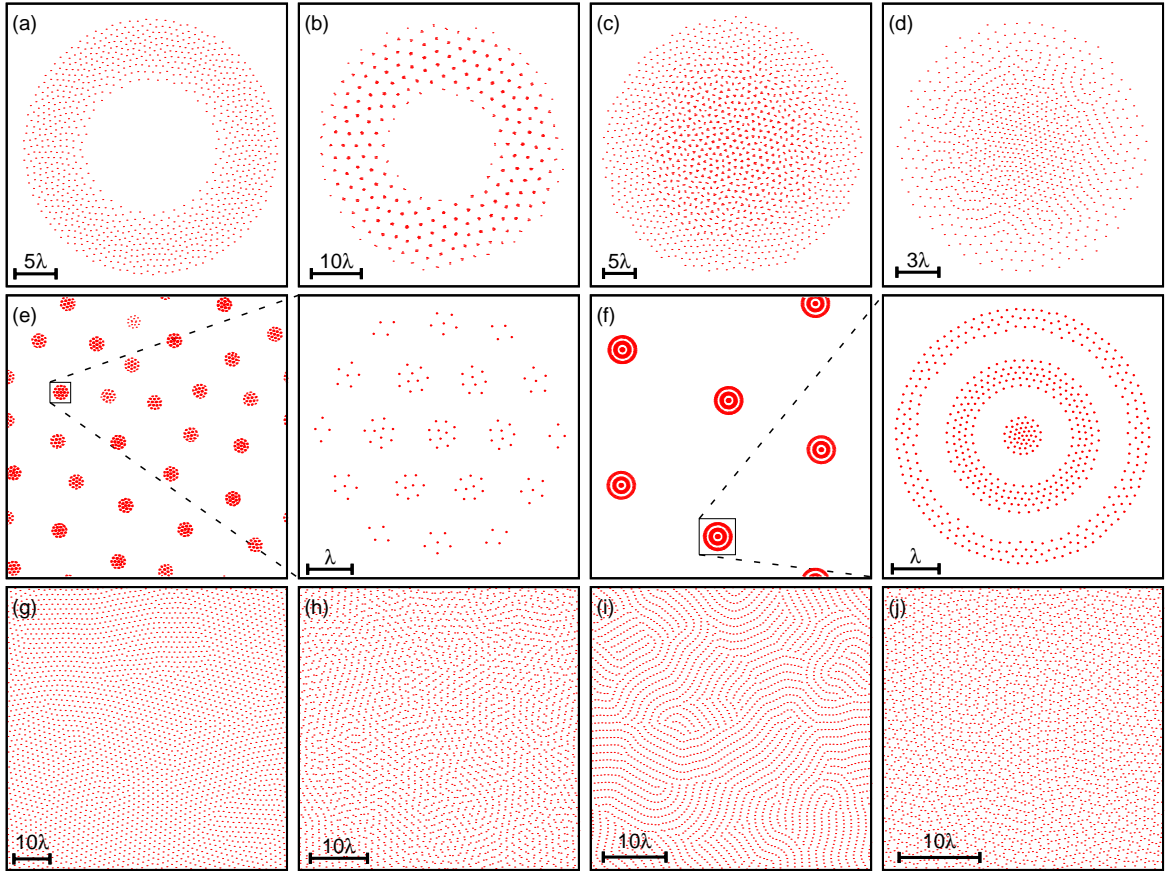}
  \caption{
    Snapshot of the final vortex configurations
    corresponding to the potentials in (a) top panel of
    \Fref{fig:potentials}(a) with $N_v=1000$ and density
    $\rho=0.044$, (b) bottom panel of \Fref{fig:potentials}(a)
    with $N_v=1000$ and density $\rho=0.025$, (c) top panel of
    \Fref{fig:potentials}(b) with $N_v=2000$ and $\rho=0.20$, (d)
    bottom panel of \Fref{fig:potentials}(b) with $N_v=1000$ and
    $\rho=0.40$, top panel of \Fref{fig:potentials}(c) with
    $N_v=3000$ and densities (e) $\rho=0.25$ and (f) $\rho=1.00$, and
    the bottom panel of \Fref{fig:potentials}(c) with $N_v=3000$
    and densities (g) $\rho=0.50$, (h) $\rho=1.25$, (i) $\rho=1.50$,
    and (j) $\rho=2.50$. For panels (a), (b), (c), and (d), the
    long-range attraction causes all of the particles to form a single
    object and we only show a close up view. The unlabeled panels are
    close-up views of panels (e) and (f), focusing on a single
    supercluster and ring, respectively. The final vortex
    configurations in panels (a)-(d) and panels (e-j) are from
    MC and LD simulations, respectively, with simulation details
    discussed in the Appendices. 
    \label{fig:snapshots}
  }
\end{figure*}

A more physically accurate potential is achieved by smoothing out the
steps in the previous potential [see \Fref{fig:schematic}(b)]. In
\Fref{fig:step}(b), we show the final MC configuration for the
smooth potential, which is a supercluster of higher radial
symmetry. The RDF is plotted in \Fref{fig:step}(d) and has nearly
identical features, although the higher symmetry of the ground state
results in smoother peaks. The conclusions that follow are (i)
multiple repulsive length scales result in a formation of hierarchical
structures and (ii) the precise form of the potential is of a lesser
importance in this example: the crucial aspect is the existence of
several length scales in the interaction.

\section{General Layered Systems}
\label{sec:gen}
To classify possible hierarchical structures we need to consider what
sorts of vortex structures exist in systems where there exist
competing interactions that are dominant at different length scales
(see \Fref{fig:potentials}). In the context of \Fref{fig:schematic},
such systems can be realized by adding layers of type-1 material or
alternating layers of clean and dirty material, while repulsive scales
are tunable by controlling e.g. layer thicknesses.

In \Fref{fig:potentials}(a), we show two potentials that both feature
a strong repulsive core surrounded by an area of attraction. Outside
the attractive shell, there is a repulsive region and an attractive
long-range coupling. At very low vortex densities, the final MC
configurations for both of these potentials are given in
\Fref{fig:snapshots}(a) and \Fref{fig:snapshots}(b), respectively. In
the first case, the particles form a cluster due to their attractive
interaction, the repulsive scale however gives this cluster a ring
shape. In the second, the combination of attractive and repulsive
scales induces clustering inside the ring (a clustered ring).

Next, we consider two potentials where the long-range attraction is
extremely weak and we vary the potential at intermediate length scales
[see \Fref{fig:potentials}(b)]. The final MC configurations for these
potentials at densities $\rho=0.2$ and $\rho=0.8$ are shown in
Figures.~\ref{fig:snapshots}(c) and \ref{fig:snapshots}(d). In the
first case, the particles form a single large supercluster with one
critical difference from the case shown on \Fref{fig:step}: the size
of the constituent small clusters is modulated by the distance to the
center of the cluster, going from a maximum of 4 vortices per cluster
in the center to a shell of single vortices at the edge. In the second
case, when the interaction in the intermediate region is modified, the
local phase of the cluster varies with distance from the center:
namely as one goes from the center of the cluster to the edge one
encounters regions corresponding to vortex lattice, vortex stripes,
and vortex voids phases. Here the long range attractive interaction
makes the vortex density gradually increase towards the center of the
cluster leading to a sequence of phases which optimize the interaction
associated with repulsive short-range scales.

The third pair of potentials [see \Fref{fig:potentials}(c)] we examine
feature a moderate repulsive core surrounded by an attractive well and
have a long-range repulsive interaction. When the well is strong, the
final vortex configurations from LD simulations~\cite{note1} are
circular superclusters at a density $\rho = 0.25$ or concentric rings
at a density of $\rho=1$ [illustrated in
Figures.~\ref{fig:snapshots}(e) and \ref{fig:snapshots}(f)]. Again the
short-range structure is determined by repulsive length
scales. However, when the attractive well is weakened significantly
more regular repulsion-dominated vortex phases appear at higher
densities [see Figures.~\ref{fig:snapshots}(g)-(j)]: a triangular
lattice, a pair vortex lattice, stripes, and voids, which is
consistent with a dominance of short-range two-scale repulsive
interactions.

\begin{figure}[tb]
  \centering
  \includegraphics[width=0.93\columnwidth]{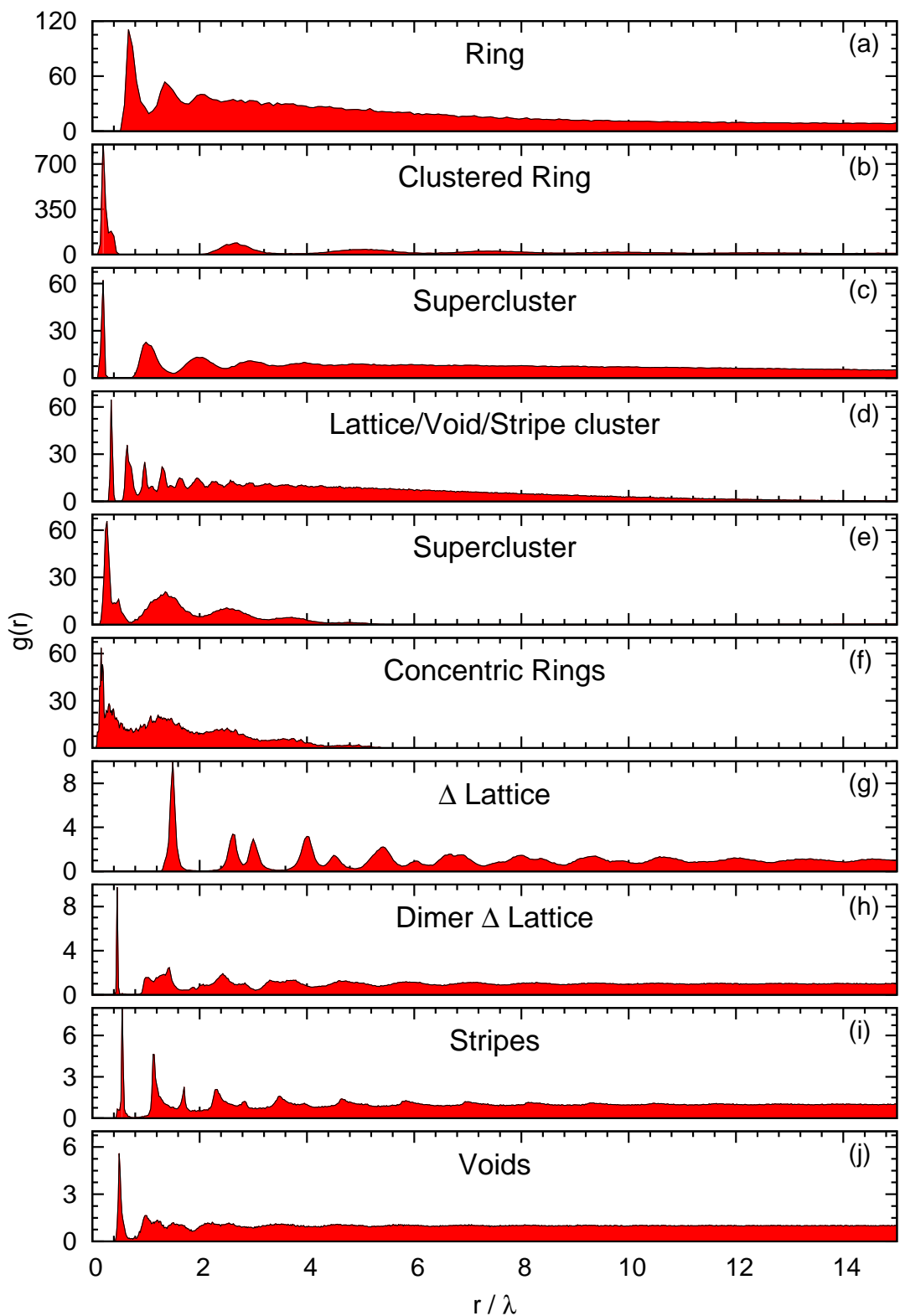}
  \caption{ Radial distribution function $g(r)$
    corresponding to the phases illustrated in
    \Fref{fig:snapshots}.
    \label{fig:rdf}
  }
\end{figure}

To better understand the structure of these phases, let us examine the
RDF for each phase, which are shown in \Fref{fig:rdf} (note that
the ordering of the panels matches the ordering of
\Fref{fig:snapshots}). For the ring phase of
\Fref{fig:snapshots}(a), $g(r)$ has three pronounced peaks
indicating the nearest-neighbor distance, next-nearest-neighbor
distance, {\it etc.} inside the ring. In between the peaks, $g(r)$
remains finite because the vortices do not form an even lattice inside
the ring. For the clustered ring phase [see
\Fref{fig:snapshots}(b)], the first peak is very pronounced and
narrow, indicative of the particles in each cluster being roughly
equidistant. The second peak characterizes nearest-neighbor distance
between clusters, and the subsequent peaks the distance between
next-nearest-neighboring clusters, etc. The long distances between the
peaks indicate that the clusters are small compared to their
separation.

The modulated supercluster of \Fref{fig:snapshots}(c) likewise has
a sharp, narrow peak in $g(r)$ representing the nearly equidistant
particle separation inside each cluster. As with the clustered ring,
the ensuing peaks represent the distances between neighboring clusters
and the broadness of the peaks is due to the variability of cluster
size throughout the structure. For the stripe/void-rich phase [see
\Fref{fig:snapshots}(d)], we observe several prominent peaks in
$g(r)$ that are largely consistent with the RDF for a triangular
lattice, with broadening of the peaks due to the mixing of phases.

The superclusters shown in \Fref{fig:snapshots}(e) possess short
range periodicity inside each cluster, resulting in two narrow peaks
in $g(r)$ which are so close together that the second peak appears as
a shoulder. The successive peaks illustrate the distance between
clusters and are broadened because of the finite size of each
cluster. Because the clusters are widely separated, there are
additional broad peaks in $g(r)$ that occur for large $r$ and describe
the supercluster separation. The RDF of the concentric ring phase
[pictured in \Fref{fig:snapshots}(f)] is remarkably similar to the
$g(r)$ for the supercluster phase
[\Fref{fig:snapshots}(e)]. Unlike the supercluster phase, $g(r)$
remains finite for $r$ smaller than the diameter of the outer ring due
to the particles spreading out evenly throughout each ring.

Finally, we discuss the radial distribution functions of more
conventionally ordered phases. The lattice phase [see
\Fref{fig:rdf}(g)] possesses much more long-range order than all
other phases considered, with peaks at $r = a,$ $\sqrt{3}a$, $2a$,
$\sqrt{7}a$, $3a$, $\ldots$, where $a$ is the nearest-neighbor
distance for a triangular lattice. The dimer lattice phase [see
\Fref{fig:rdf}(h)] has a clearly defined peak describing the dimer
separation. Because the pairs prefer to line up end-to-end, the peaks
that would describe the triangular lattice are broadened significantly
and the long-range order cannot be observed in $g(r)$. The stripe
phase [see \Fref{fig:rdf}(i)] has several regularly spaced peaks
coinciding perfectly with the separation of particles along each
stripe. Here, broadening occurs due to both bending of the stripes and
the presence of other stripes. The void phase [see
\Fref{fig:rdf}(j)] only has short-range periodicity, as evidenced
by the pronounced features for small $r$.

\section{Summary}
\label{sec:summary}
In this paper, we presented MC and LD simulations of vortex states for
a model of vortex-vortex interactions in general layered
superconductor-insulator-superconducting structures, made of different
superconducting layers. The vortices are subject to interactions with
multiple length scales. We have shown that these layered systems have
an unusual magnetic response: vortex supercluster structures, which
can consist of clusters of clusters, rings, clusters in a ring, or
have coexistence of stripes, voids, and lattice phases. This can
provide an experimental tool to deduce information about vortex
interactions from observation of their ordering in real experiments.
Besides that it indicates that one can use layered superconducting
structures for the realization and study of rich and unique pattern
forming systems. More generally, our results indicate that systems
with additional characteristic length scales may exhibit more
complicated hierarchies in structure formation. In such systems,
increasing the number of repulsive or attractive length scales should
result in multiple hierarchy orderings, yielding ``fractal crystal''
phases in the limit of a large number of length scales.

In further studies we plan to quantitatively determine the multi-scale
inter-vortex interactions by solving coupled Ginzburg-Landau and
Maxwell equations for superconductor-insulator multi-layers.
 
\ack
We thank M. Touminen and C. Santangelo for discussions. This work was
supported by NSF Award No. DMR-0955902 (C.N.V., Q.W. and E.B.) by Knut
and Alice Wallenberg Foundation through the Royal Swedish Academy of
Sciences (E.B.), Swedish Research Council (E.B. AND K.S.). The
computations were partially performed on resources provided by the
Swedish National Infrastructure for Computing (SNIC) at National
Supercomputer Center at Linkoping, Sweden.

\appendix
\section{Methods Summary}
\label{sec:methods}
In this paper we have utilized both MC and LD simulations to obtain
the vortex structure formations at $T = 0$.

\paragraph*{Monte Carlo.}
The vortex structure formations are obtained using the Metropolis MC
algorithm, where at each MC iteration a randomly chosen vortex is
displaced a distance $(0,d]$ chosen at random, and MC moves are
accepted or rejected according to the Metropolis MC scheme. The
potential energy is calculated with a sharp cut off of the interaction
potential at half the box length. The presented results are snapshots
at zero temperature after at least $10^3$ sweeps from an random
initial configuration, where temperature was incrementally lowered to
zero. The step length $d$ was typically set to the box length.

\paragraph*{Langevin Dynamics.}
The dynamics of a vortex are described by the overdamped Langevin
equation of motion
\begin{equation}
  \eta \deriv{{\bf r}_i}{t} = {\bf F}_i^{vv} + {\bf F}_i^T,
\end{equation}
where $\eta$ is the damping constant, ${\bf r}_i$ is the position of
the $i$-th vortex, ${\bf F}_{i}^{vv} = -\nabla V_{ij}$ is the
inter-vortex force, and ${\bf F}_i^T$ is the stochastic thermal
force. The simulation is performed with periodic boundary conditions
and the vortex-vortex interaction is cut off
smoothly~\cite{Fangohr2000,Fangohr2001}. The equation of motion is
integrated by an Euler scheme with a reduced time step of $\Delta t =
0.005$. We measure length in units of
$\lambda=\unit{200}{\angstrom}$. In all cases, the initial
configuration of the vortices is randomly distributed, and the
temperature is incrementally lowered to zero, with at least $2 \times
10^5$ integration steps at each temperature. For 3000 vortices, each
time step takes $\sim\unit{2.2}{\second}$ on a single AMD processor
with a clock speed of $\unit{2.2}{\giga\hertz}$.

%\subsection*{Radial Distribution Function}
\paragraph*{Radial Distribution Function.}
The phases are analyzed by means of the radial distribution
function (RDF)~\cite{Hansen1986}
\begin{equation}
  g(r) = \frac{1}{2 \pi r \Delta r \rho N} \sum_{i=1}^N n_i(r,\Delta r),
\end{equation}
where $n_i(r,\Delta r)$ is the number of particles in the shell
surrounding the $i$th particle with radius $r$ and thickness $\Delta
r$. For small distances, $g(r) \to 0$. Meanwhile, for large distances
the radial distribution function must approach unity, i.e. $g(r) \to
1$ as $r \to \infty$.

\section{Potentials}
\label{sec:potentials}
In order to understand the competition between interactions acting at
different length scales, we first consider a core-softened hard-sphere
model with multiple shoulders
\begin{equation}
  V(r) = \left\lbrace
    \begin{array}{ccc}
    \infty & &r < \sigma_h\\
    \varepsilon_1 & &\sigma_h < r < \sigma_1\\
    \varepsilon_2 & &\sigma_1 < r < \sigma_2\\
    0 & &r > \sigma_2
  \end{array}\right.,
  \label{eq:step}
\end{equation}
where $\sigma_h$ is the hard-core radius, $\sigma_1$ and $\sigma_2$
are the shoulder radii, and $\varepsilon_1$ and $\varepsilon_2$ are
the shoulder heights (see Fig.~1(a) for a sketch of a vortex in this
potential). In this work, we performed MC simulations with $N_v =
2000$ vortices, various densities and parameters $\sigma_h = 0.25$,
$\sigma_1 = 1.75$, $\sigma_2 = 10$, $\varepsilon_1 = 2.25$, and
$\varepsilon_2 = 0.8$. The ground state of this pair potential for a
wide range of densities is the supercluster phase illustrated in
Fig.~2(a).

The key feature responsible for different structure formations is the
number of repulsive/attractive length scales and their relative
values. The precise shape of the potential is less important. A more
physical realization of this potential can be obtained by smoothing
out the steps [see Fig.~1(b)]. This smoothed potential can be modeled
by
\begin{equation}
  V(r) = \left(K_0(r) + e^{-(r-1)^2} + 1\right) e^{-1/(10-r)},
  \label{eq:p1b}
\end{equation}
where $K_0$ is a modified Bessel function of the second kind, $r$ is
the separation between particles, and $V(r)=0$ for $r\ge10$. In all
cases studied, we found the ground state to be similar to the ground
state of \Eref{eq:step}.

As discussed in the main text, general layered systems have competing
interactions that are not necessarily repulsive. Attractive parts can
arise from core-core interactions in layers of type-1 material.
Figure~3 of the main text illustrates several scenarios considered in
this study. The potential form of Fig.~3(a) is modeled by
\begin{equation}
  \frac{V(r)}{V_0} = \frac{e^{-r/\lambda}}{r/\lambda} - c_1
  e^{-\alpha_1(r/-\beta_1)^2} + c_2 e^{-\alpha_2(r/a-\beta_2)^2},
  \label{eq:p3a}
\end{equation}
where $c_1$, $\alpha_1$, $\beta_1$, $c_2$, $\alpha_2$, and $\beta_2$
are coefficients, $V_0$ sets the unit of energy, and $\lambda$ is the
characteristic length scale of magnetic field localization. In what
follows, we set $\lambda=1$ as the unit of length. The values of these
coefficients corresponding to the top and bottom panels of Fig.~3(a)
are $c_1=1.0$, $\alpha_1=0.005$, $\beta_1=20$, $c_2=100$,
$\alpha_2=50$, $\beta_2=0.1$ and $c_1=1$, $\alpha_1=0.01$,
$\beta_1=30$, $c_2=10$, $\alpha_2=1.0$, $\beta_2=1.0$, respectively.

The potential form corresponding to Fig.~3(b) is very similar to
\Eref{eq:p3a}, 
\begin{equation}
  \frac{V(r)}{V_0} = \frac{e^{-r/\lambda}}{r/\lambda} - c_1
  e^{-\alpha_1(r/-\beta_1)} + c_2 e^{-\alpha_2(r/-\beta_2)^2}.
  \label{eq:p3b}
\end{equation}
Here the coefficients for the top and bottom panels are $c_1=0.1$,
$\alpha_1=0.1$, $\beta_2=0.0$, $c_2=5.0$, $\alpha_2=10$, $\beta_2=0.5$
and $c_1=1.0$, $\alpha_1=1.0$, $\beta_2=0.0$, $c_2=0.6$,
$\alpha_2=100$, $\beta_2=0.5$, respectively.

The potential for Fig.~3(c) is
\begin{equation}
    \frac{V(r)}{V_0} =  e^{-r / \lambda} -
    c_2 e^{-r / \xi}
    + c_3 \frac{\lambda \lbrace\tanh[\alpha(r -
      \beta)]+1\rbrace}{r + \delta},
  \label{eq:p3c}
\end{equation}
It models a short-range exponential repulsion, an intermediate-ranged
exponential attraction, and a long-range power law term resulting from
stray fields. For the third term, the onset of the long-range force is
mediated by the parameters $a$, $b$, and $\delta$, which ensure that
the dominant short-range force is the first term. For both the top and
bottom panels, we consider $c_3 = 0.1$, $\alpha=2.5$, $\beta=0.5$, and
$\delta=0.1$. The remaining parameters are $c_2 = 0.9$ and $\xi =
1.80$ for the top panel and $c_2 = 0.5$ and $\xi = 1.85$ for the bottom
panel.

\bibliographystyle{iopart-num}
\bibliography{references}

\end{document}